\documentclass[preprint,aps,preprint,pra]{revtex4}
\usepackage{graphicx}

\begin{document}
\title{Hole doped Dirac states in silicene by biaxial tensile strain}
\author{T. P. Kaloni, Y. C. Cheng, and U. Schwingenschl\"ogl}
\email{udo.schwingenschlogl@kaust.edu.sa,+966(0)544700080}
\affiliation{PSE Division, KAUST, Thuwal 23955-6900, Kingdom of Saudi Arabia}

\begin{abstract}
The effects of biaxial tensile strain on the structure, electronic states, and mechanical
properties of silicene are studied by ab-initio calculations. Our results show that up to
5\% strain the Dirac cone remains essentially at the Fermi level, while higher strain
induces hole doping because of weakening of the Si$-$Si bonds. We demonstrate that the
silicene lattice is stable up to 17\% strain. It is noted that the buckling first decreases
with the strain (up to 10\%) and then increases again, which is accompanied 
by a band gap variation. We also calculate the Gr\"uneisen parameter and demonstrate
a strain dependence similar to that of graphene.
\end{abstract}

\maketitle

\section{Introduction}
Silicene is a two dimensional buckled material which is closely related to graphene. It has been proposed as a
potential candidate for overcoming the limitations of graphene because of stronger intrinsic spin orbit coupling 
(4 meV in silicene and $1.3\cdot10^{-3}$ meV in graphene \cite{cheng1}). Silicene first has been reported 
to be stable by Takeda and Shiraishi \cite{takeda}. Though C and Si belong to the same group of the 
periodic table, Si has a larger ionic radius, which promotes $sp^3$ hybridization. Theoretical studies predict 
that free standing silicene has a stable two-dimensional buckled honeycomb structure \cite{ciraci,olle}, where the buckling is 
due to the mixture of $sp^2$ and $sp^3$ hybridizations. The magnitude of the buckling is 
$\sim$ 0.45 \AA, which opens an electrically tunable band gap \cite{falko,Ni}, whereas the induced band gap due to the 
intrinsic spin orbit coupling amounts to 1.55 meV \cite{yao}. The charge carriers behave like massless Dirac 
fermions in the $\pi$ and $\pi^*$ bands, which form Dirac cones at the Fermi level at the K and K$'$ points. 
The electronic properties of silicene and its derivatives have been studied in much detail by density functional 
theory calculations \cite{houssa,Bechstedt,kang,Wang}. In particular, it has been reported that the 
lattice is sensitive to the carrier concentration but still stable in a wide range of doping \cite{cheng}.

Experimentally, growth of silicene and its derivatives has been reported for metallic substrates like Ag and ZrB$_2$ 
\cite{padova,vogt,ozaki}. Silicene on a ZrB$_2$ thin film shows an asymmetric buckling due to strong interaction with 
the substrate, which increases the band gap. As in general accurate measurements of materials properties are problematic
on substrates, it is desirable to achieve free standing silicene. 
However, this first requires the growth on appropriate substrates that make it possible to separate the silicene sheet. 
For the growth of silicene on any kind of substrate, the effect of strain is 
crucial to be understood. In this work, we focus on this topic using first-principles calculations. 
We apply strain up to 20\% and calculate the corresponding band structure to evaluate the dependence of the 
induced doping on the strength of the biaxial tensile strain. Furthermore, we study the phonon spectrum to address the 
stability of the system and calculate the Gr\"uneisen parameter.

\section{Computational details}
We have carried out calculations using density functional theory in the generalized gradient approximation \cite{paolo}. 
The van der Waals interaction \cite{grime,jmc} is taken into account in order to correctly describe the geometry. 
The calculations are performed with a plane wave cutoff energy of 816 eV. Moreover, a Monkhorst-Pack $16\times16\times1$ k-mesh is 
employed for optimizing the crystal structure and calculating the phonon spectrum, whereas a $24\times24\times1$ k-mesh is 
used for the density of states (DOS) in order to achieve higher resolution. The atomic positions are relaxed until 
an energy convergence of 10$^{-9}$ eV and a force convergence of $4\cdot10^{-4}$ eV/\AA\ are reached. We use an interlayer 
spacing of 16 \AA\ to avoid artifacts of the periodic boundary conditions. The magnitude of the biaxial tensile strain is 
defined as $\varepsilon = \frac{(a-a_0)}{a_0} \times 100\%$, where $a$ and $a_0 = 3.86$ \AA\ are the lattice 
parameters of the strained and unstrained silicene, respectively.

\begin{figure*}[ht]
\includegraphics[width=0.5\textwidth,clip]{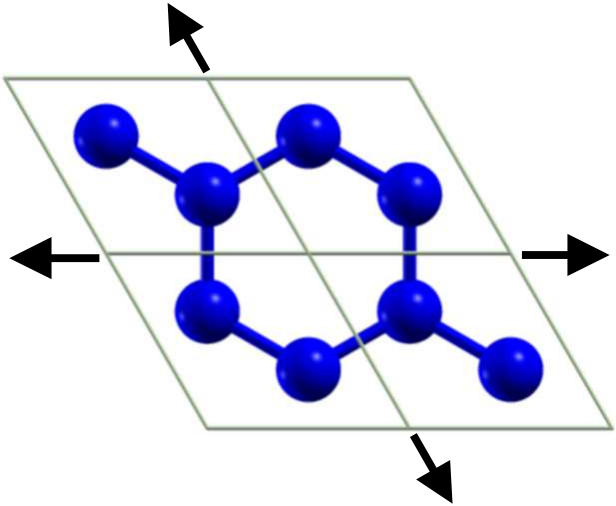}
\caption{Crystal structure of silicene under consideration. The arrows indicate the direction of the biaxial tensile strain.}
\end{figure*}

\section{Results and discussion}

For graphene it has been demonstrated that 5 to 10\% strain can be achieved without much
efforts \cite{Andresa}. The existing reports confirm this makes the system five times more
reactive and H atoms are bound much stronger than in pristine graphene \cite{Andresa}.
Since a similar enhancement of H storage by strain can be expected for silicene, we 
study in the following the effect of strain on the electronic and mechanical properties.
A top view of the crystal structure under consideration is shown in Fig.\ 1. For unstrained
silicene we obtain a lattice parameter of $a = 3.89$ \AA\ and a buckling of 0.45 \AA,
consistent with previously reported data \cite{ciraci,cheng}. In a first step, we address 
the dependence of the force on the applied strain, see the results in Fig.\ 2. The force
increases monotonically with the strain up to a strain of 17\% and decreases thereafter,
which indicates that silicene is stable up to 17\% strain. The stability limit will be
addressed in more detail via the phonon spectrum in the following section.

\begin{figure*}[ht]
\includegraphics[width=0.5\textwidth,clip]{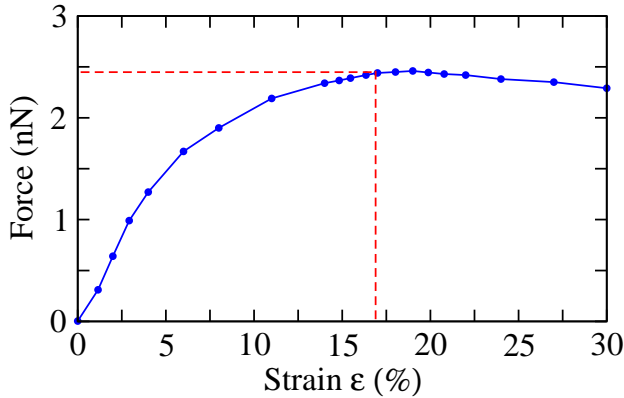}
\caption{Variation of the force as a function of the applied biaxial tensile strain.}
\end{figure*}

The band gap of 2 meV in unstrained silicene becomes smaller for increasing strain. Since
strain weakens the internal electric field (by reducing the magnitude of the buckling) the
spin orbit coupling and thus the induced band gap are reduced. The Si$-$Si bond length is
found to grow with the strain monotonically, which explains why the buckling decreases. 
Surprisingly, the buckling starts to increase again when the strain exceeds 10\%. 
For example, unstrained silicene has a Si$-$Si bond length of 2.28 \AA\ and buckling of
0.46 \AA. For 5\% strain these values change to 2.37 \AA\ and 0.32 \AA, and for 17\%
strain to 2.47 \AA\ and 0.30 \AA. The variation of the Si$-$Si bond length and buckling
under strain are addressed in Fig.\ 3(a) and (b), respectively.

\begin{figure*}[t]
\includegraphics[width=0.5\textwidth,clip]{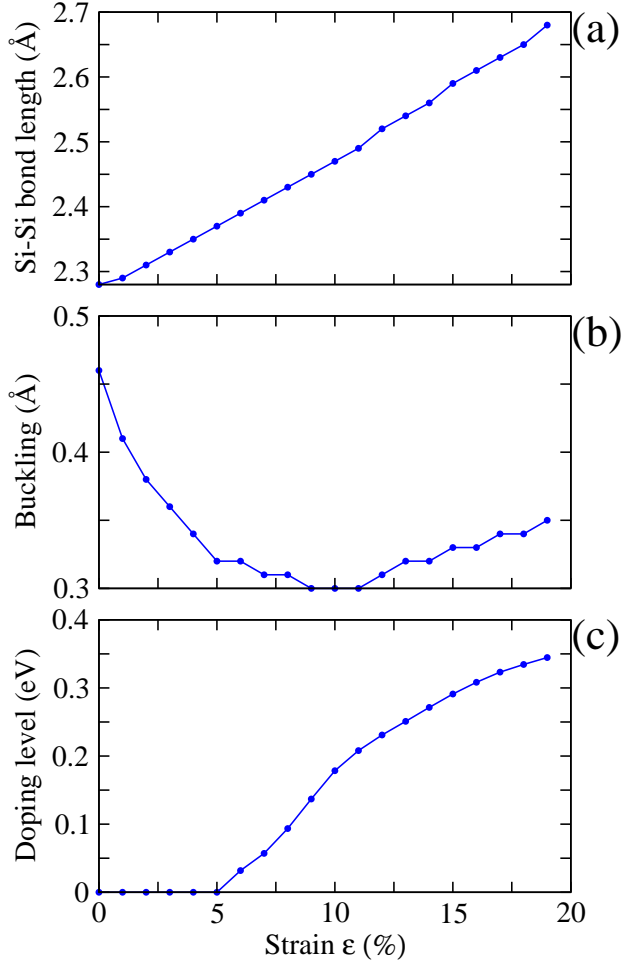}
\caption{Variation of (a) the Si$-$Si bond length, (b) the buckling, and (c) the doping level under biaxial tensile strain.}
\end{figure*}

The variation of the doping level (defined as the shift of the Dirac cone with respect to
the Fermi level) under strain is addressed in Fig.\ 3(c). It is well known that unstrained
silicene is a semimetal, where the $p_z$ and $p_z^*$ orbitals give rise to $\pi$ and $\pi^*$
bands forming Dirac cones at the K and K$'$ points, see Fig.\ 4(a). The calculated band 
structure shows that the Dirac cone lies at the Fermi level upto a strain of 5\% with a
2 meV band gap due to intrinsic spin orbit coupling. For higher strain the conduction band
at the $\Gamma$-point shifts towards the Fermi level, consistent with Ref.\ \cite{liu}.
At a strain of 7\% it slightly crosses the Fermi level, which shifts the Dirac cone 
above the Fermi level by $\sim$ 0.06 eV, inducing hole doping, see Fig.\ 3(c). The doping
is enhanced for increasing strain, since the conduction band minimum at the $\Gamma$-point
shifts further downwards and becomes more and more occupied (with an increasing DOS at
the Fermi level). The main reason for hole doping in silicene under strain is this
downshift and the consequent occupation of the band at the $\Gamma$-point. It is a
consequence of the weakening of the bonds due to the increasing Si$-$Si bond length.
Another ingredient is a reduction of the hybridization between the $s$ and $p$ orbitals,
which in fact are occupied by 1.18 and 2.76 electrons in unstrained silicene, respectively,
but by 1.33 and 2.63 electrons for 10\% strain.

\begin{figure*}[t]
\includegraphics[width=0.8\textwidth,clip]{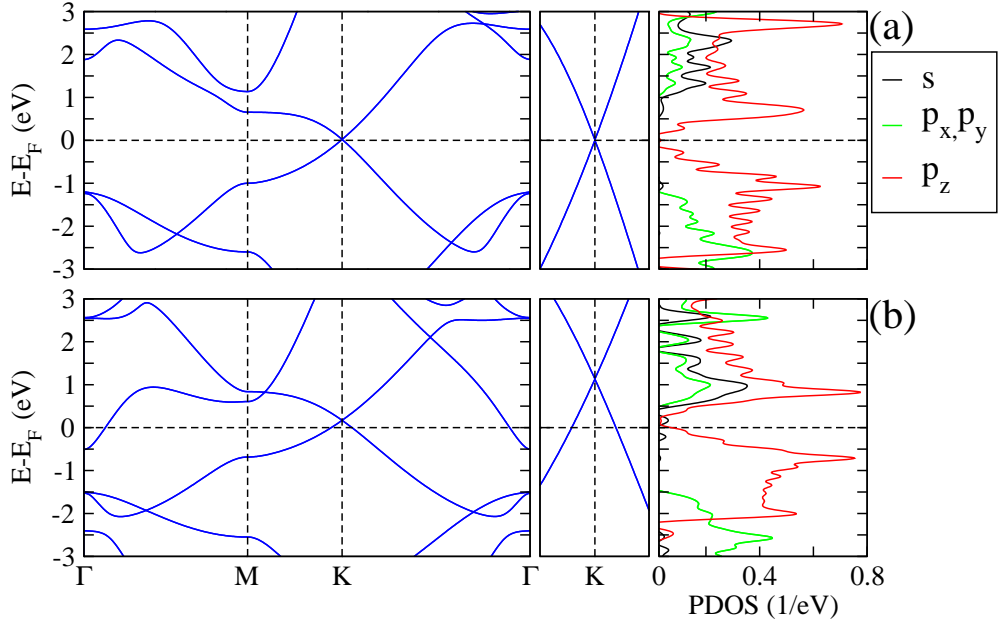}
\caption{Electronic band structure with corresponding partial DOSs for (a) unstrained and (b) 10\% biaxially tensile strained silicene.}
\end{figure*} 

At 10\% strain the Dirac point lies at 0.18 eV, see Fig.\ 4(b). We note that the $\pi$ and
$\pi^*$ bands are due to the $p_z$ orbitals with minute contributions from the $p_x$ and
$p_y$ orbitals, as expected, see the projected DOSs. For higher strain the conduction band
minimum shifts further to lower energy and the Dirac cone accordingly to higher 
energy. It reaches 1.0 eV with the Dirac point at 0.34 eV for a strain of 20\%. This behavior
is different from graphene despite the quantitatively similar band structure, because the
Si$-$Si bonds are more flexible than the C$-$C bonds. In contrast to silicene, graphene
does not show significant changes in the electronic structure in the presence of strain,
resulting a zero band gap semiconductor up to a huge strain of 30\% \cite{son}. As a result,
doping cannot be achieved in graphene by strain. 

We now discuss the phonon spectrum of silicene without strain and under strain of 5\%, 10\%,
15\%, 20\%, and 25\%. Without strain the optical phonon frequencies are found to be $\sim$
33\% smaller than in graphene \cite{cheng}, which is understood by the smaller force
constant and weaker Si$-$Si bonds. In fact, the Si$-$Si bond length of 2.28 \AA\ is 
37\% larger than the C$-$C bond length. In Fig.\ 5 we address the phonon band structure,
where we focus on the highest branches at the $\Gamma$-point (G mode) and the K-point
(D mode). The calculated phonon frequencies at the $\Gamma$ and K-points are 550 cm$^{-1}$
and 545 cm$^{-1}$, respectively, which agree well with previous theoretical results
\cite{ciraci,cheng}. A significant modification of the phonon frequencies is observed for
strained silicene. For a strain of 5\% the G and D mode frequencies amount to 460 cm$^{-1}$
and 386 cm$^{-1}$, respectively, reflecting the weakening of the Si$-$Si bond under strain. 
Increase of the strain to 10\% (17\%) results in phonon frequencies of 372 cm$^{-1}$
(296 cm$^{-1}$) for the G mode and 272 cm$^{-1}$ (187 cm$^{-1}$) for the D mode. We still
have positive frequencies along the $\Gamma$-K direction and, hence, a stable lattice. An
instability comes into the picture when the strain increases beyond 17\%. At 20\% strain
we find a frequency of $-5$ cm$^{-1}$ and at 25\% strain, see Fig.\ 5(c), the lattice is
strongly instable. Importantly, no splitting of the G mode for increasing strain is
observed in our calculations in contrast to graphene \cite{udo}.

\begin{figure*}[t]
\includegraphics[width=0.5\textwidth,clip]{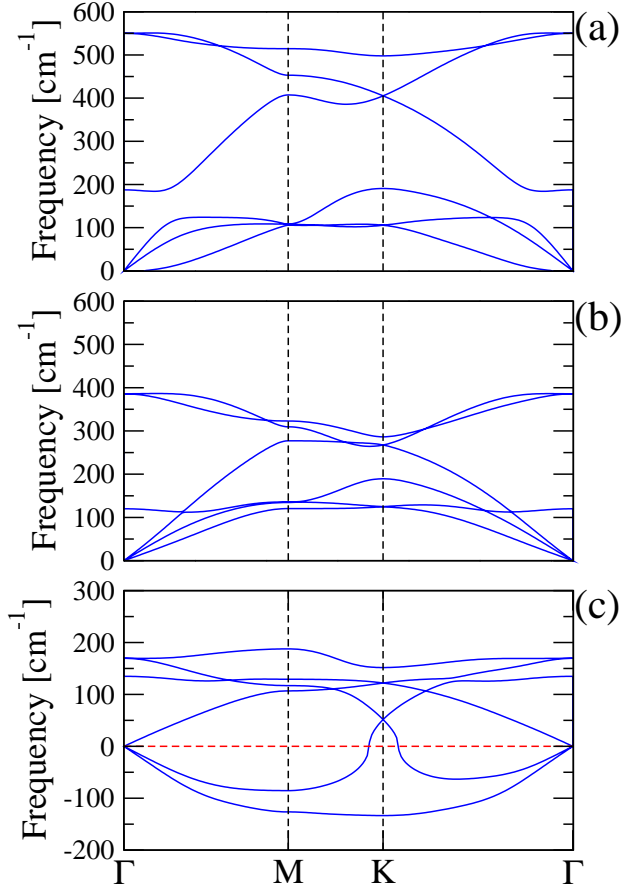}
\caption{Phonon frequencies for (a) unstrained, (b) 10\% biaxially tensile strained, and (c) 25\% biaxially tensile strained silicene.}
\end{figure*}

The Gr\"uneisen parameter is an important quantity to describe strained materials as it
measures the rate of phonon mode softening or hardening and, thus, determines the
thermomechanical properties. The Gr\"uneisen parameter for the G mode is given by
\begin{displaymath}
\gamma_G=-\Delta\omega_G/2\omega_G^0\varepsilon,
\end{displaymath}
where $\Delta\omega_G$ is the difference in the frequency with and without strain and
$\omega_G^0$ is the frequency of the G mode in unstrained silicene. A significant variation
of the Gr\"uneisen parameter between 1.64 and 1.42 for strain between 5 and 25 \% is found,
see Table I. These values are close to the experimental and theoretically values for graphene
\cite{Mohiuddin,ding,udo,Remi}. While the experimentally reported Gr\"uneisen parameters for
graphene are not consistent due to substrate effects, there are no experimental data
available for silicene for comparison. We find that the Gr\"uneisen parameter first decreases
with growing strain due to the reduced buckling of the two Si sublattices but increases
again for higher strain as also the buckling increases. This behavior is fundamentally
different from graphene, which is not subject to buckling. An experimental confirmation
of our observations by Raman spectroscopy would be desirable.

\begin{table*}[ht]
\begin{tabular}{|c|c|c|}
\hline
$\varepsilon$ (\%) & $\Delta\omega_G$ (cm$^{-1}$) &${\gamma_G}$ \\
\hline
5          & 460                & 1.64 \\
\hline
10          & 372               & 1.62 \\
\hline
15          & 296               & 1.54 \\
\hline
20          & 246               & 1.34 \\
\hline
25          & 160               & 1.42 \\
\hline
\end{tabular}
\caption{Strain, frequency shift of the G mode, and Gr\"uneisen parameter of the G mode.}
\end{table*}

\section{Conclusion}
In conclusion, we have used density functional theory to study the effect of biaxial tensile
strain on the structure, electronic properties, and phonon modes of silicene. Our calculations
demonstrate that up to 5\% strain the Dirac cone remains essentially at the Fermi level but
starts to shift to higher energy for higher strain. Therefore, strain can be used in
silicene, in contrast to graphene, to induce hole doping. The different behavior of the two 
compounds, despite their close stuctural similarity, can be explained in terms of bonding
and changes in the hybridizations. Strain results in a weakening of the Si$-$Si bonds. As
a consequence, an electronic band at the $\Gamma$-point of the Brillouin zone shifts to
lower energy and becomes partially occupied, which in turn leads to a depopulation of the
Dirac cone. The buckling is found to decrease with increasing strain up to 10\% but starts
to increase again thereafter. Accordingly, the calculated Gr\"uneisen parameter behaves
differently than in graphene as the latter is not subject to buckling. Positive phonon 
frequencies up to a strain of 17\% indicate lattices stability in this regime, whereas the
lattice becomes instable at higher strain.

\end{document}